\documentclass[journal=jacsat,manuscript=article]{achemso}
\usepackage{indentfirst}
\usepackage[version=3]{mhchem} % Formula subscripts using \ce{}
\usepackage{xcolor} % For \textcolor command
\author{André A. A. Silva}
\affiliation{Brazilian Synchrotron Light Laboratory (LNLS), Brazilian Center for Research in Energy and Materials (CNPEM), Campinas, SP, 13083-100, Brazil}
\alsoaffiliation{Gleb Wataghin Physics Institute (IFGW), University of Campinas(Unicamp), 13083-859, Campinas, SP, Brazil}
\altaffiliation{Equivalent contribution}

\author{Eduardo D. Stefanato}
\affiliation{Instituto de Física, Universidade de São Paulo, São Paulo, SP, 05508-090, Brazil}
\altaffiliation{Equivalent contribution}

\author{Nicolas M. Kawahala}
\affiliation{Instituto de Física, Universidade de São Paulo, São Paulo, SP, 05508-090, Brazil}

\author{Flávio H. Feres}
\affiliation{Brazilian Synchrotron Light Laboratory (LNLS), Brazilian Center for Research in Energy and Materials (CNPEM), Campinas, SP, 13083-100, Brazil}

\author{João Vítor T. P. Vital}
\affiliation{Physics Department, Instituto de Ciências Exatas, Universidade Federal de Minas Gerais, Belo Horizonte, MG, 31270-901, Brazil}

\author{Bernardo R. A. Neves}
\affiliation{Physics Department, Instituto de Ciências Exatas, Universidade Federal de Minas Gerais, Belo Horizonte, MG, 31270-901, Brazil}

\author{Jan Plutnar}
\affiliation{Dept. of Inorganic Chemistry, University of Chemistry and Technology Prague, Technicka 5, 166 28 Prague 6, Czech Republic}

\author{Zdenek Sofer}
\affiliation{Dept. of Inorganic Chemistry, University of Chemistry and Technology Prague, Technicka 5, 166 28 Prague 6, Czech Republic}

\author{Ana C. F. Brito}
\affiliation{Brazilian Synchrotron Light Laboratory (LNLS), Brazilian Center for Research in Energy and Materials (CNPEM), Campinas, SP, 13083-100, Brazil}

\author{Felix G.G. Hernandez}
\affiliation{Instituto de Física, Universidade de São Paulo, São Paulo, SP, 05508-090, Brazil}
\email{felixggh@if.usp.br}

\author{Raul O. Freitas}
\affiliation{Brazilian Synchrotron Light Laboratory (LNLS), Brazilian Center for Research in Energy and Materials (CNPEM), Campinas, SP, 13083-100, Brazil}
\email{raul.freitas@lnls.br}

\title[An \textsf{achemso} demo]
  {Magnetic permeability and zone-folded phonons in layered NiPS\textsubscript{3}}

\abbreviations{IR,NMR,UV}

\begin{document}


\begin{abstract}
\noindent Nickel phosphorus trisulfide (NiPS$_{3}$) stands out as a Mott insulator exhibiting XY-type antiferromagnetic order. In this work, we investigate the electrodynamic response of single-crystaline NiPS$_{3}$ using Terahertz Time-Domain Spectroscopy (THz-TDS) as a function of temperature. In cases where the magnetic permeability cannot be approximated to unity, conventional THz transmission analysis is often restricted to the complex refractive index, hindering the distinction between magnetic and dielectric contributions. To overcome this limitation, we demonstrate an extraction methodology capable of decouple the magnetic permeability from the electric permittivity. Assuming that the dielectric contribution remains invariant below the Néel temperature (T$_{N}$), we used data from the paramagnetic phase to isolate the intrinsic magnetic component at 13~K. The angular and thermal dependence of the spectra revealed a magnon mode near 1 THz (for $0^{\circ}$ orientation) and, distinctively, a zone-folded phonon activated by symmetry breaking (at $90^{\circ}$). The isolated magnetic permeability was successfully modeled using Drude-Lorentz oscillators, enabling the extraction of fundamental spin dynamics parameters in this material.
\end{abstract}

\section{Introduction}

Transition metal phosphorus trichalcogenides (MPX$_3$, where M is a transition metal, P is phosphorus, and X is a chalcogen) have recently attracted renewed interest as a promising class of van der Waals (vdW) semiconductors owing to their correlated insulating and antiferromagnetic properties. This coexistence of phenomena makes nickel phosphorus trisulfide (NiPS$_3$) an attractive platform for exploring low-dimensional magnetism and for developing spintronic and optoeletronic applications.\cite{Woongki2024,Su2024,Afanasiev2021,Yan2021,Wang2020,Jenjeti2018,Wang2018,Burch2018,Park2016,Lee2016,Huang2017,Joy1992} NiPS$_3$ is a correlated Mott insulator that exhibits XY-type antiferromagnetic (AFM) order below the Néel temperature ($\text{T}_N = 155$ K).\cite{Afanasiev2021,Kang2020,Lee2016,Belvin2021} The robustness of this order and the strong correlation between spin and lattice degrees of freedom make it an ideal platform for exploring the dynamics of collective excitations and their fundamental interactions.\cite{Allington2025,Toyoda2024,Woongki2024,Su2024,Afanasiev2021,Belvin2021,Wang2020}

In the low-energy spectral range, particularly in the terahertz (THz) frequencies, the electrodynamic response of $\text{NiPS}_3$ is governed by magnon and phonon modes whose properties depend on the polarization direction.\cite{Woongki2024,Su2024,Afanasiev2021,Belvin2021,Toyoda2024,Allington2025} Below $\text{T}_N$, the antiferromagnetic order induces a translational symmetry break that enables Brillouin zone folding, hence, allowing vibrational modes at the Brillouin-zone center that are forbidden in the paramagnetic phase. Recent studies\cite{Woongki2024,Su2024,Afanasiev2021,Belvin2021,Toyoda2024,Allington2025} point to the existence of active modes at frequencies near 1 THz, whose origin, whether purely magnetic or arising from a zone-folded phonon, these vibrational modes requires further evaluation of their dispersion behavior for a deeper knowledge on their fundamentals.

Terahertz time-domain spectroscopy (THz-TDS) is a well-established technique for probing the absorption response of materials in the spectral range from hundreds of GHz to a few THz. Because it measures the transmitted electric-field waveform in the time domain, both the amplitude and phase of the radiation–matter interaction can be determined simultaneously, enabling the direct extraction of complex optical response functions, such $\varepsilon(\omega) = \varepsilon_1(\omega) + i\varepsilon_2(\omega)$ and $\mu(\omega) = \mu_1(\omega) + i\mu_2(\omega)$. The technique becomes even more powerful when combined with a cryostat, allowing measurements across a wide temperature range and enabling the sample to be driven through phase transitions in its phase diagram. In the case of $\text{NiPS}_3$, temperature control across the Néel temperature ($\text{T}_N$) provides direct access to the evolution of magnetic excitations and spin–lattice interactions, making it possible to track the emergence or suppression of collective modes associated with the antiferromagnetic ordering. 

However, conventional analyses of THz transmission spectra in magnetic materials typically rely on the complex refractive index, $\tilde{n}(\omega) = n(\omega) + i\kappa(\omega)$, which does not readily isolate the purely magnetic contribution of magnon modes \cite{nvemec2006independent, pronin2009phase,Neu2018}. In general, the refractive index depends simultaneously on the complex permittivity and permeability, $\tilde{n}=\sqrt{{\varepsilon} {\mu}}$, and therefore cannot uniquely distinguish electric and magnetic resonances when both responses are relevant \cite{nvemec2006independent, Papari2025}. Némec et al. demonstrated that the independent determination of the complex refractive index and wave impedance enables the retrieval of the dielectric permittivity and magnetic permeability in bulk materials using THz-TDS \cite{nvemec2006independent}. More recently, this approach has been extended to thin-film systems, enabling the simultaneous extraction of $\varepsilon$ and $\mu$ from THz measurements \cite{Papari2025}. In contrast to these approaches, which require independent spectroscopic information from transmission and reflection measurements or from multiple internal echoes, here we exploit the temperature dependence of the THz response to disentangle the dielectric and magnetic contributions using transmission measurements of a bulk crystal.

In anisotropic systems such as $\text{NiPS}_3$, where magnetic resonances occur only for specific polarization, determining the frequency dispersion of the $\mu(\omega)$ is mandatory for accurately modeling electromagnetic wave propagation for in connection to those excitations. In this work, we demonstrate an extraction methodology to decouple the magnetic permeability from the electric permittivity in the optical response of $\text{NiPS}_3$. We also report the observation of a zone-folded phonon in the THz range, previously detected only by Raman Spectroscopy.\cite{Woongki2024,Su2024} By exploiting the temperature and polarization dependence of the response, we assume that above $\text{T}_N$, where the material is in the paramagnetic phase and magnetic resonances are absent, the magnetic permeability approaches unity, $\mu(\omega) \approx 1$, enabling the independent determination of the electric permittivity, $\varepsilon(\omega)$. Assuming the dielectric contribution remains unchanged below $\text{T}_N$, we used room-temperature data to isolate the magnetic component in the spectra measured at 13~K. The resulting permeability was modeled using Drude-Lorentz oscillators,\cite{Grishunin2018, Kawahala2023Coatings, Stefanato2025} enabling the extraction of the fundamental parameters of the magnon mode. The temperature dependence of the modes further allowed us to distinguish magnetic excitations (magnons) from non-magnetic excitations (phonons).

%This is a paragraph of text to fill the introduction of the
%demonstration file.  The demonstration file %attempts to show the
%modifications of the standard \LaTeX\ macros %that are implemented by
%the \textsf{achemso} class.  These are mainly %concerned with content,
%as opposed to appearance.

\newpage
\section{Results and Discussion}

\begin{figure}[h!]
  \centering
  \includegraphics[width=0.5\textwidth]{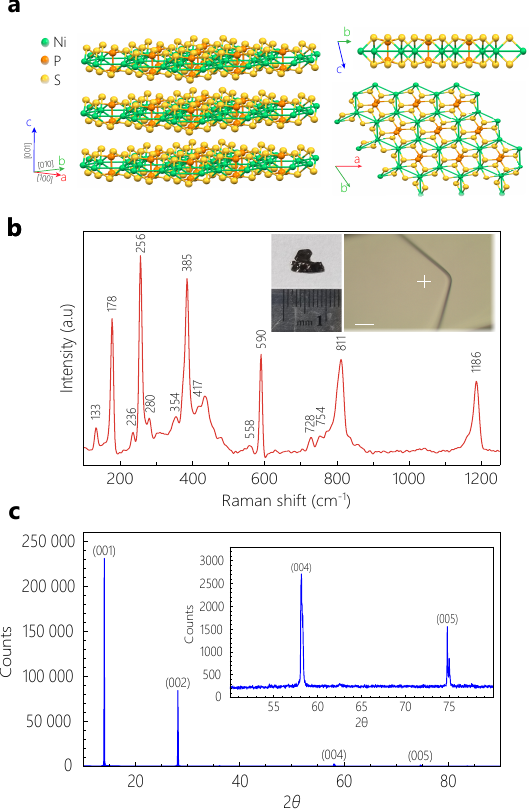}
  \caption{\textbf{Structural characterization of the $\text{NiPS}_3$ crystal.} \textbf{(a)} Schematic representation of the $\text{NiPS}_3$ crystal structure, belonging to the monoclinic system with space group C2/m. The illustration displays projections along the crystallographic planes formed by the $b$-$c$ axes (highlighting the van der Waals layer stacking along the $c$-axis) and the $a$-$b$ axes (highlighting the distorted hexagonal arrangement of metal ions in the basal plane). Spheres represent Ni (green), P (orange), and S (yellow) atoms. \textbf{(b)} Raman spectrum of the $\text{NiPS}_3$ bulk sample at room temperature. The observed vibrational modes confirm the structural integrity of the material. The inset shows an optical microscopy image of the $\text{NiPS}_3$ flake. \textbf{(c)} X-Ray Diffraction (XRD) pattern of the sample obtained using Cu - $K {\alpha}$ radiation. The sharp and well-defined diffraction peaks, indexed according to the monoclinic C2/m phase, attest to the high crystallinity and phase purity of the single crystal.}
  \label{Fig1}
\end{figure}

$\text{NiPS}_3$ is a layered member of the transition-metal thio-
phosphate family, in which $Ni^{2+}$ ions form a two-dimensional honeycomb lattice interconnected by
[$P_{2}S_{6}$] units,\cite{Woongki2024,Belvin2021,Toyoda2024} as illustrated in Figure 1(a). The layers are weakly bound by van der
Waals interactions, configuring a quasi-two-dimensional system that allows mechanical
exfoliation. To verify the crystalline phase of the samples investigated here, we performed Raman spectroscopy and X-ray diffraction (XRD), as shown in Figures~\ref{Fig1}(b, c). Figure~\ref{Fig1}(b) shows the Raman peaks of a bulk $\text{NiPS}_3$ flake (thickness $\approx$ 450 $\mu$m), measured with excitation at 633 nm. The spectrum agrees well with previous reports,\cite{Bernasconi1988,Kuo2016} and the eight Raman-active phonon modes are assigned accordingly. XRD measurements further confirms a monoclinic crystal structure with space group $C2/m$, in agreement with the literature.\cite{Du2016,Jenjeti2018,Yan2021}

\begin{figure}[h!]
  \centering
  \includegraphics[width=1.0\textwidth]{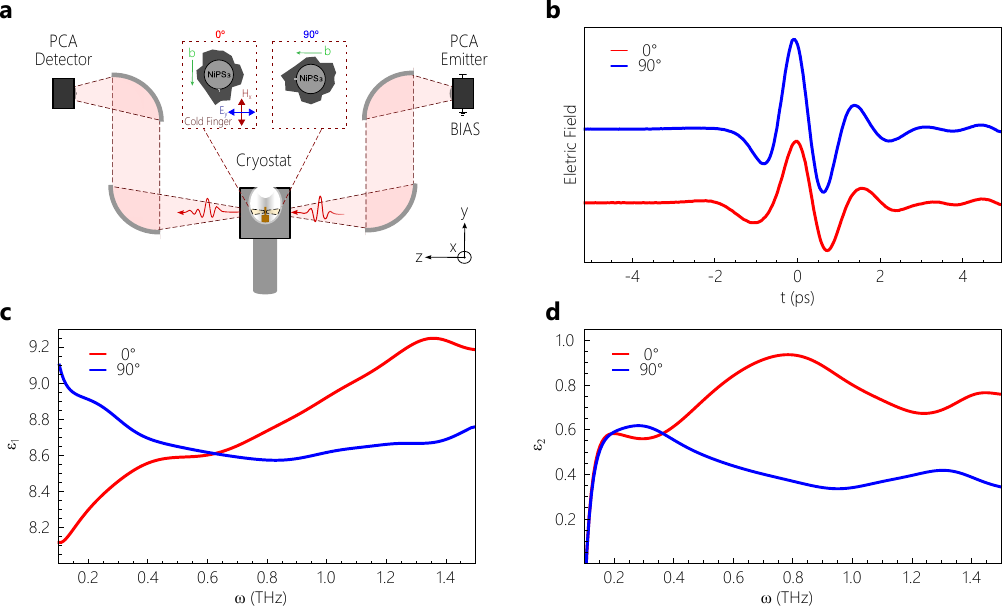}
  \caption{\textbf{Experimental setup and anisotropic dielectric characterization of NiPS$_3$ at room temperature.} (a) Schematic illustration of Terahertz Time-Domain Spectroscopy (THz-TDS) in transmission geometry. The system is driven by a femtosecond laser and employs photoconductive antennas (PCAs) for the emission and coherent detection of radiation. The THz beam passes through the sample positioned inside a cryostat (for thermal control) or a standard sample holder, with the entire optical path enclosed in a purge box to eliminate water vapor absorption. (b) Time-domain waveforms of the THz pulse transmitted through the $\text{NiPS}_3$ crystal at room temperature (T = 300~K), measured for two orthogonal crystal axis orientations ($0^{\circ}$ and $90^{\circ}$) relative to the angle that the crystallographic \textit{b}-axis makes with the magnetic field THz in a clockwise direction. (c, d) Complex electric permittivity spectra extracted using a THz electric field transmission model within the thick sample approximation, assuming the absence of magnetic modes at room temperature ($\mu(\omega) \approx 1$), i.e., negligible magnetic contribution. The angular dependence ($0^{\circ}$ vs. $90^{\circ}$) confirms the in-plane dielectric anisotropy of the crystal in the paramagnetic phase, requiring characterization for each orientation.}
  \label{Fig2}
\end{figure}

The optical response of NiPS$_3$ was characterized using a THz-TDS system \cite{Kawahala2025, Dias2026}, as illustrated in Figure~\eqref{Fig2}(a), along two distinct directions defined as follows, the $0^{\circ}$ orientation corresponds to the configuration where the THz magnetic field is aligned parallel to the crystallographic \textit{b}-axis. The $90^{\circ}$ orientation was obtained by rotating the sample 90° clockwise relative to the polarization of the magnetic field THz. In other words, the polarization of the incident THz electric and magnetic fields remained fixed, and the relative orientation was controlled solely by rotating the sample. The experimental setup relies on the coherent generation and detection of THz radiation pulses using photoconductive antennas (PCAs) driven by a femtosecond laser. To ensure spectral accuracy and to eliminate water vapor absorption lines, we enclosed the entire optical path of the THz beam within a purge box (relative humidity $<$ 5\%).

Initially, time-domain profiles of the electric field transmitted through the bulk sample at room temperature were obtained for two distinct crystal orientations, as presented in Figure~\ref{Fig2}(b). By Fourier processing these signals and employing the THz transmission equations within the thick sample approximation, Eqs.~\eqref{eq1} and \eqref{eq2}, the complex electric permittivity, $\varepsilon(\omega)$, was extracted. Figures~\ref{Fig2}(c,d) display the real and imaginary parts of this parameter, respectively, highlighting anisotropy in the material's dielectric response. This pronounced in-plane dieletric anisotropy in the paramagnetic phase underscores the necessity of precise crystallographic alignment when investigating these van der Waals antiferromagnets. Establishing this room-teperature baselise, $\varepsilon(\omega)$, is a mandatory for isolating any emergent magnetic phenomena at lower temperatures, given that the magnetic and dielectric responses are intrinsically coupled within the measured refractive index. 

\begin{figure}[h!]
  \centering
  \includegraphics[width=.8\textwidth]{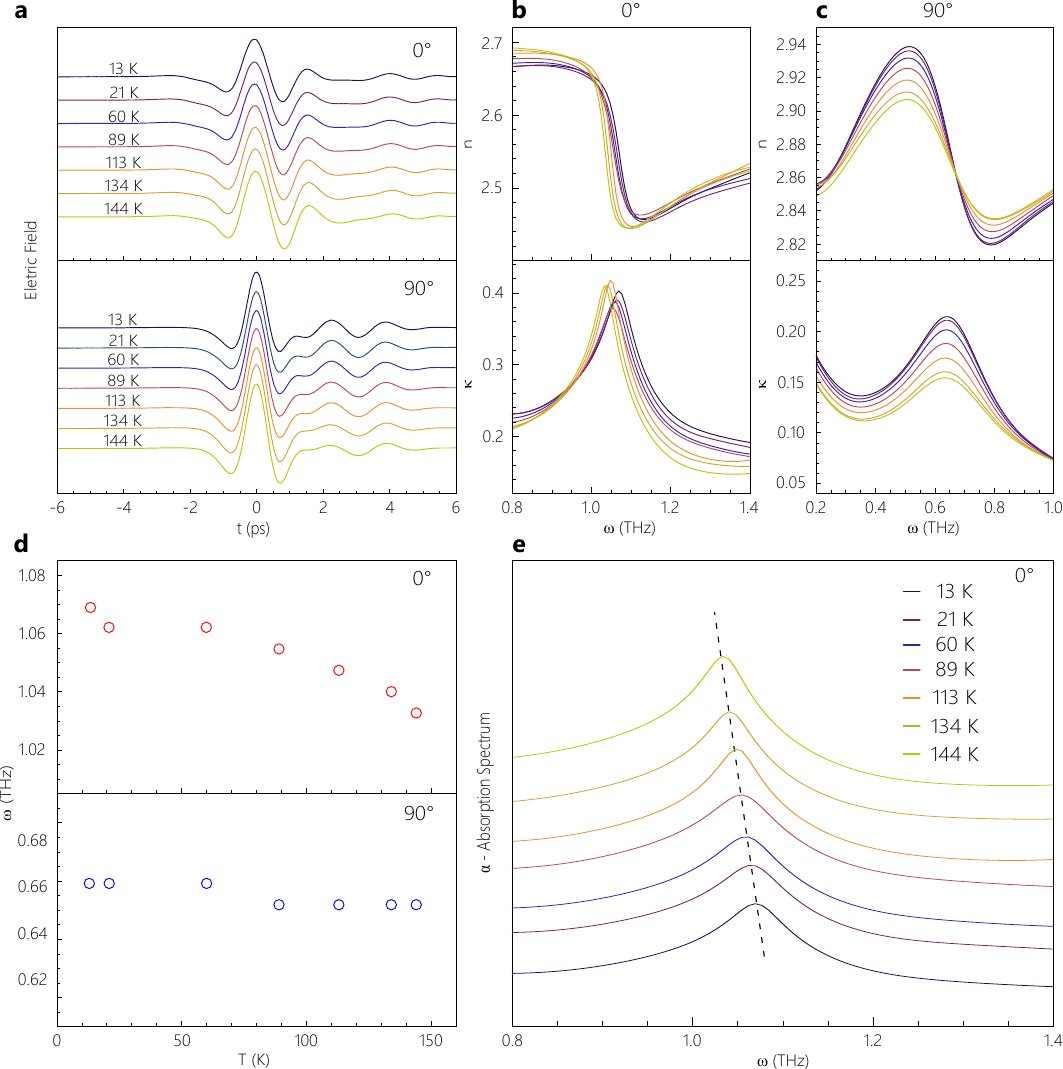}
  \caption{\textbf{Thermal evolution of the Terahertz response and mode dispersion in $\text{NiPS}_3$ for different orientations.} (a) Time-domain profiles of THz pulses transmitted through the sample across the temperature range from 144~K to 13~K for the $0^{\circ}$ and $90^{\circ}$ orientations. Curves are vertically offset for clarity. (b, c) Real and imaginary parts, of the temperature-dependent complex refractive index for the $0^{\circ}$ and $90^{\circ}$ orientations. For the $0^{\circ}$ orientation, a resonance near 1.0 THz is observed, exhibiting temperature-dependent dispersion of its center frequency. For the $90^{\circ}$ orientation, a resonance near 0.6 THz is present, in this case, we do not observe dispersion of the center frequency with temperature, but rather an increase in resonance amplitude upon cooling. (d) Dispersion plots (mode center frequency vs. temperature) extracted from the absorption spectrum fo the $0^{\circ}$ orientation and from the imaginary part of the complex refractive index for the $90^{\circ}$ orientation. (e) Absorption coefficient spectrum for the $0^{\circ}$ orientation, highlighting the emergence of the mode upon cooling and its temperature-dependent dispersion. The black dashed line highlights the blueshift in the magnon mode frequency as a function of temperature shown in (d). Curves are vertically offset for clarity.}
  \label{Fig3}
\end{figure}

To investigate the mode dynamics, temperature-dependent measurements were performed in the ordered phase, from 144~K to 13~K. Since the T$_N$ of $\text{NiPS}_3$ is well established in the literature to be around 155~K, restricting our low-temperature analysis to this range ensures that the material is unambiguously in the antiferromagnetic phase, which validates the assumption that the magnetic permeability deviates from unity, allowing the application of our extraction methodology. Figure~\ref{Fig3} summarizes the optical parameters for the different orientations. In the time-domain waveforms presented in Figure~\ref{Fig3}(a), a qualitative change is observed upon the cooling down of the material. For the $0^{\circ}$ orientation, the emergence of a resonant mode near 1 THz is evident. The dispersion analysis of this orientation, shown in Figure~\ref{Fig3}(d), shows a blue shift as the temperature decreases, which is expected for magnon mode.\cite{Afanasiev2021,Woongki2024,Belvin2021} In contrast, for the $90^{\circ}$ orientation, the spectral response reveals distinct dynamics and resonance frequency. Here the observed mode dispersion exhibits a thermal behavior different from that of the previous orientation.

The dispersion analysis for the $0^\circ$ orientation, shown in Figure~\ref{Fig3}(d), reveals a clear blueshift of the mode's center frequency as the temperature decreases. This behavior is an intrinsic characteristic of magnon modes,\cite{Afanasiev2021,Woongki2024,Belvin2021} reflecting the consolidation of long-range magnetic order in $\text{NiPS}_3$. Upon cooling the sample below T$_N$, the reduction in thermal fluctuations promotes a stiffening of the internal exchange field and an increase in magnetic anisotropy. Since these interactions act as the primary restoring forces for the spin system, their enhancement raises the energy barrier required to excite collective precessions, resulting in the experimentally observed blueshift of the magnon resonance towards $\sim 1.07$ THz at 13~K. In stark contrast, the resonance observed at $90^\circ$ does not exhibit this characteristic magnetic dispersion. Its thermal evolution, where the center frequency remains almost constant around $\sim 0.65$ THz while the amplitude increases upon cooling, is consistent with a zone-folded phonon, activated by the translational symmetry breaking of the antiferromagnetic ordering, which renders a mode previously inactive in the paramagnetic phase active in the
antiferromagnetic phase.\cite{Su2024,Woongki2024} Although previous studies have employed Raman scattering to detect these lattice vibrations, our findings demonstrate that THz-TDS provides a direct and complementary probe for these spin-coupled quasiparticles in the low energy regime.

\begin{figure}[h!]
  \centering
  \includegraphics[width=.75\textwidth]{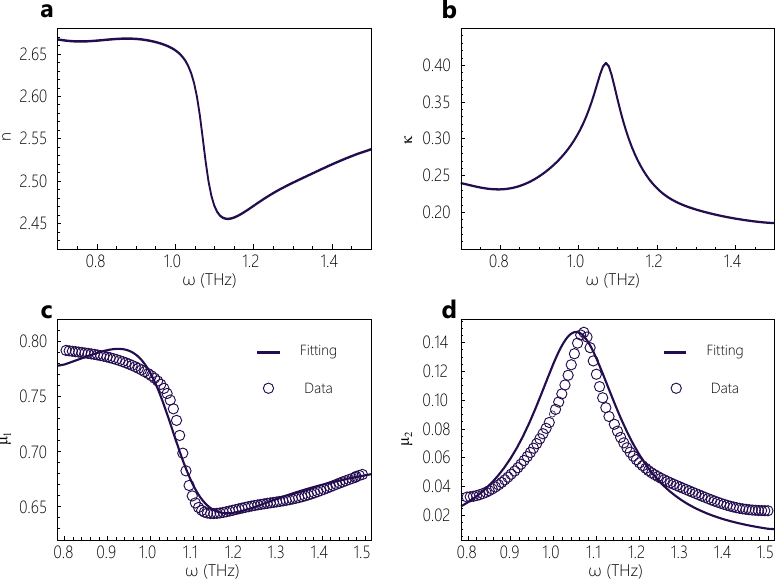}
  \caption{\textbf{Extraction of the complex magnetic permeability of the magnon mode at 13~K (0° orientation).} (a, b) Components of the measured complex refractive index: (a) Real and (b) Imaginary part. These spectra represent the total optical response, containing mixed contributions from electric permittivity and magnetic permeability. The solid line represent the experimental data. (c, d) Isolated complex magnetic permeability obtained via the extraction methodology established in this work: (c) Real part and (d) Imaginary part. The data (dots) were calculated by removing the background dielectric contribution obtained in the paramagnetic phase at 300~K. The solid line represents the theoretical fit using the standard Drude-Lorentz form, confirming the Lorentzian nature of the magnetic resonance.}
  \label{Fig4}
\end{figure}

The application of the extraction methodology to decouple, described in the Methods section, enables the isolation of the intrinsic magnetic response of the material. Figures~\ref{Fig4}(a, b) present the real and imaginary parts, respectively, of the complex refractive index measured at 13~K for the orientation $0^{\circ}$. In principle, the dielectric and magnetic responses are coupled within the measured refractive index. To extract the complex magnetic permeability, we assume that the electric permittivity does not undergo abrupt changes upon decreasing temperature for this orientation exhibiting only a single magnetic mode within the spectral range approached here, such that $\varepsilon(13~\text{K}) \approx \varepsilon(300~\text{K})$. Using the room-temperature data as the dielectric background, the magnetic permeability was calculated according to Eq.~\eqref{eq3}.

The results of this decoupling are shown in Figures~\ref{Fig4}(c, d). The imaginary part of the permeability, Figure~\ref{Fig4}(d), reveals a distinct absorption peak. Simultaneously, the real part, Figure~\ref{Fig4}(c), exhibits a dispersive curve characteristic of an isolated resonance. Finally, the extracted experimental data were fitted using a Drude-Lorentz model, Eq.~\eqref{eq4}, enabling the determination of the central frequency, $\omega_0 = (1.0598 \pm 0.0018)$ THz, linewidth, $\Delta\omega = (0.2524 \pm 0.0071)$ THz, and oscillator strength, $\Delta\mu = (3.497 \pm 0.094) \times 10^{-2}$, for the magnon mode. The successful isolation of $\mu(\omega)$ validates the assumption regarding the thermal stability of the dielectric background. By attributing the resonant spectral changes exclusively to spin dynamics, we circumvet the limitations of standard analyses relying solely on the refractive index, enabling a more fundamental analysis from the perpective of light-matter interaction. The good agreement between the extracted data and the Drude-Lorentz profile not only confirms the efficacy of our extraction methodology, but also yields intrinsic magnetic parameters. The extraction of these parameters is crucial for accurately modeling the electrodynamic response of NiPS$_3$, laying the groundwork for its integration into future technological applications. 

\section{Conclusions}

We established the extraction methodology by distinguishing between the dielectric and magnetic responses across the phase transition. At room temperature, above the transition temperature, $\text{NiPS}_3$ behaves as a paramagnetic medium where the magnetic contribution is assumed to be negligible, $\mu(\omega) \approx 1$. Consequently, the response observed in the transmission spectra exclusively reflects the material's complex electric permittivity, given that $\tilde{n}^2 = \sqrt{\varepsilon(\omega)\mu(\omega)}$. The central premise of our analysis is the thermal stability of this dielectric background. By assuming that fluctuations in $\varepsilon(\omega)$  are negligible upon cooling, remaining virtually constant, we can attribute the resonant spectral changes observed at 13~K solely to spin dynamics. Thus, we employ the dielectric parameter obtained at 300~K as a reference to extract the magnetic permeability $\mu(\omega)$ in the ordered phase.

The distinction between the $0^{\circ}$ and $90^{\circ}$ orientations reveals the physical nature of the observed resonances. The mode at $0^{\circ}$, with its characteristic blue shift at low temperatures, exhibits behavior typical of a magnon mode near 1 THz, corroborating previous literature reports\cite{Woongki2024,Afanasiev2021,Belvin2021}. In contrast, the mode observed in the $90^{\circ}$ orientation, Figure~\ref{Fig3}(d), does not display the thermal dispersion of a magnetic resonances. Its behavior is consistent with that of a zone-folded phonon\cite{Woongki2024,Su2024}, activated by translational symmetry breaking in the antiferromagnetic phase. It is noteworthy that, while previous literature restricted the observation of these modes to Raman Spectroscopy, our data demonstrates that THz-TDS is a robust complementary tool for investigating these quasiparticles coupled to the magnetic order.

Finally, the efficacy of the extraction methodology is evidenced by the quality of the fit using a Drude-Lorentz model on the isolated magnetic permeability curves, Figures~\ref{Fig4}(c, d). This enabled the precise determination of fundamental parameters, such as the central frequency, linewidth, and oscillator strength of the magnon mode, validating the methodology for separating dielectric and magnetic contributions in the THz range.

\section{Methods}

%\textbf{Transmission Model}.%
The samples were mounted on the cold finger of a cryostat equipped with terahertz-transparent windows and aligned normal to the incident THz beam. Off-axis parabolic mirrors were used to focus the THz radiation onto the sample and to collect the transmitted signal. The transmitted THz electric field was detected using an optically gated photoconductive antenna, enabling direct time-domain measurements of the waveform with appropriated windowing \cite{Marulanda2025}.

To probe anisotropic and polarization-dependent responses, the NiPS$_3$ crystals were measured at rotation angles of 0$^\circ$ and 90$^\circ$ in the plane relative to the polarization of the incident THz magnetic field. The time-domain signals were Fourier processed to obtain the complex frequency-dependent transmission. Reference measurements acquired under identical conditions were used to determine the complex experimental transmission coefficient $\tilde{\mathcal{T}}_{\exp}(\omega)$.

The complex refractive index was extracted directly from $\tilde{\mathcal{T}}_{\exp}(\omega)$ using a standard phase--amplitude analysis for a plane-parallel slab of thickness $d$. The real part of the refractive index $n(\omega)$ and the extinction coefficient $\kappa(\omega)$ were obtained as\cite{Neu2018, Kawahala2023Coatings}

\begin{equation}
    \begin{cases}
    n(\omega) = 1 + \frac{c}{\omega d} \arg\left[\tilde{\mathcal{T}}_{\exp}(\omega)\right], \\
    \kappa(\omega) = -\frac{c}{\omega d} \ln \left[ \frac{(n(\omega) + 1)^2}{4n(\omega)} \left|\tilde{\mathcal{T}}_{\exp}(\omega)\right| \right],
    \end{cases}
    \label{eq1}
\end{equation}
where $c$ is the speed of light in vacuum and $\omega$ is the frequency. The complex dielectric function, $\varepsilon(\omega)$, was then obtained from the direct relation

\begin{equation}
    \varepsilon(\omega) = \tilde{n}^2(\omega)
    = n^2 - \kappa^2 + 2in\kappa.
    \label{eq2}
\end{equation}
In this analysis, we initially assume that the magnetic permeability is close to unity, a condition assumed to be considered valid at room temperature. However, below the transition temperature, the magnetic permeability deviates from unity, and in this regime the complex refractive index alone is insufficient to distinguish between the contributions of permeability and permittivity. 

However, based on the hypothesis that the electric permittivity does not undergo significant variations with decreasing temperature, it is possible to decouple the permeability using the room-temperature permittivity as a reference. Therefore, by applying the constitutive relation connecting the complex refractive index, electric permittivity, and magnetic permeability, we derive a system of equations for the real and imaginary parts of the permeability, $\mu(\omega)$, whose solutions are given by,
\begin{equation}
    \begin{cases}
    \mu_1(\omega) = \frac{2\varepsilon_2(\omega)n(\omega)\kappa(\omega)+\varepsilon_1(\omega)(n^2(\omega) - \kappa^2(\omega))}{\varepsilon_1^2(\omega) + \varepsilon_2^2(\omega)}, \\
    \mu_2(\omega) = \frac{2n(\omega)\kappa(\omega)}{\varepsilon_1(\omega)} - \frac{\varepsilon_2(\omega)}{\varepsilon_1(\omega)}\left(\frac{2\varepsilon_2(\omega)n(\omega)\kappa(\omega) - \varepsilon_1(\omega)(n^2(\omega) - \kappa^2(\omega))}{\varepsilon_1^2(\omega) + \varepsilon_2^2(\omega)}\right),
    \end{cases}
    \label{eq3}
\end{equation}
where $\varepsilon_j\,\,(j = 1, 2)$ corresponds to the value measured at room temperature, $n$ and $\kappa$ denote the values measured at the temperature of interest (below the phase transition). Finally, the magnon mode response was modeled using a Drude-Lorentz profile, given by,
\begin{equation}
    \mu(\omega) = \mu_{\infty} + \frac{\Delta\mu \omega_0^2}{\omega_o^2 - \omega^2 + i\omega\Delta\omega}.
    \label{eq4}
\end{equation}
The resulting fit, presented in Figures~\ref{Fig4}(c, d), demonstrates very good agreement with the experimental data.

\textbf{THz-TDS}. Temperature-dependent THz-TDS measurements were performed in a transmission geometry using a conventional broadband setup. Single-crystal NiPS$_3$ were placed in the sample holder with an aperture within the diffraction limits \cite{Dias2026} and investigated over a temperature range from 300~K down to 13~K to probe low-energy excitations in the terahertz frequency range. Broadband THz pulses were generated by a biased photoconductive antenna excited by infrared pulses from a mode-locked Ti:sapphire laser operating at a central wavelength of 780~nm, with a pulse duration of approximately 130~fs and a repetition rate of 76~MHz.

\indent \textbf{Synthesis of $\text{NiPS}_3$ crystals.} NiPS$_3$ was made by direct reaction from elements in quartz ampoule. Nickel (-100 mesh, 99.99\%, Alfa Aesar) sulfur (99.9999\%, 2-6mm, Wuhan Xinrong New Materials, Co, China) and phosphorus (99.9999\%, 2-6mm, Wuhan Xinrong New Materials Co, China) were placed in quartz ampoule (50c250 mm) in stochiometric ratio corresponding to 30g of NiPS$_3$ together with  1 at.\% excess of sulfur and phosphorus together with 0.7g of iodine (99.9\%, granules, Fisher Scientific, UK) and melt sealed under high vacuum ($<$1x10-3 Pa using oil diffusion pump and LN2 trap) using oxygen-hydrogen welding torch. Ampoules were placed in a muffle furnace and gradually heated on 450°C for 25 hours, on 500°C for 50 hours and finally on 600°C for 50 hours. Homogenized reacted $\text{NiPS}_3$ ampoules were placed in a two zone horizontal furnace for CVT crystal growth. First the growth zone was heated to 750 °C and the source zone was at 600 °C for 2 days. Subsequently thermal gradient was reversed and source zone was heated to 750 °C and growth zone on 650 °C. Finally, the ampoule was cooled to room temperature and crystals were collected in an argon-filled glovebox.

\textbf{Sample Characterization}. The diffraction measurements were performed using a Bruker D8 Advance ECO system with Cu-$K{\alpha}$ ($1.54178$ \r{A}) radiation. The Raman Spectroscopy experiments were performed using a confocal WITec Alpha 300RA Raman microscope in the backscattering configuration, with a 633 nm (1.96 eV) laser, a 100× objective (NA = 0.9), and a laser power of 0.5 mW. Spectra were collected with 5 accumulations of 60 s using a 600 grooves/mm grating. All measurements were performed on bulk NiPS$_3$ samples at room temperature.

\begin{acknowledgement}

All authors thank the Brazilian Synchrotron Light Laboratory (LNLS) and Brazilian Nanotechnology National Laboratory (LNNano), part of the Brazilian Centre for Research in Energy and Materials (CNPEM), a private non-profit organization under the supervision of the Brazilian Ministry for Science, Technology, and Innovations (MCTI), for sample preparation (LAM proposal 20242805) and DRX measurements (LNNano proposal 20251949). The authors would like to acknowledge the LCPNano at the Federal University of Minas Gerais (UFMG) for providing the equipment and technical support for the Raman spectroscopy experiments. Activities within the GCTI-THz/USP group were supported by the São Paulo Research Foundation (FAPESP) under Grant Nos. 2021/12470-8, 2023/04245-0, and 2023/16742-8, and by the National Council for Scientific and Technological Development (CNPq) under Grant No. 306550/2023-7. AAAS acknowledges financial support from FAPESP (Grant No. 2023/09372-0). EDS acknowledges financial support from FAPESP (Grant No. 2025/02029-3). ACFB acknowledges financial support from FAPESP through the Research, Innovation and Dissemination Center for Molecular Engineering for Advanced Materials – CEMol (Grant CEPID Nos. 2024/00989-7 and 2025/27155-1). FHF acknowledges financial support from FAPESP (Grant No. 2023/09839-5). RF Acknowledges financial support from FAPESP (Grant No. 2019/14017-9) and CNPq (Grant No. 300197/2025-0). Z.S. was supported by ERC-CZ program (project LL2101) from Ministry of Education Youth and Sports (MEYS).

\end{acknowledgement}


\bibliography{achemso-demo}

\end{document}